\documentclass[11pt]{article}
\usepackage{epsfig}

\catcode`@=12
\newcommand{\be}{\begin{equation}}
\newcommand{\ee}{\end{equation}}
\newcommand{\bea}{\begin{eqnarray}}
\newcommand{\eea}{\end{eqnarray}}
\def\r{\rightarrow}

\def\s12{\sin\theta_{12}}
\def\s23{\sin\theta_{23}}
\def\s13{\sin\theta_{13}}
\def\ts12{\theta_{12}}
\def\ta23{\theta_{23}}
\def\t13{\theta_{13}}
\def\e{\epsilon}

\makeatletter
\@addtoreset{equation}{section}

\makeatother
\title{\LARGE\bf $\mu\tau$ symmetry, tribimaximal mixing and four zero neutrino Yukawa textures}
\author{{\bf Biswajit Adhikary$^{\rm a, b}$\footnote{biswajit.adhikary@saha.ac.in}
     , Ambar Ghosal$^{\rm a}$\footnote{ambar.ghosal@saha.ac.in} and 
Probir Roy$^{\rm a}$\footnote{DAE Raja Ramanna Fellow, probir.roy@saha.ac.in}}\\
  a) Saha Institute of Nuclear Physics, 1/AF Bidhan
        Nagar, Kolkata 700064, India \\
  b)Department of Physics, Gurudas College,
Narkeldanga, Kolkata-700054, India}
\date{}
\begin{document}
\maketitle
\thispagestyle{empty}
\begin{abstract}
\noindent
Within the type-I seesaw framework with three heavy right chiral neutrinos and in the basis where 
the latter and the charged leptons are mass diagonal, a near $\mu\tau$ symmetry in the neutrino sector is strongly suggested by the neutrino oscillation data. 
There is further evidence for a close to the tribimaximal mixing pattern which 
subsumes $\mu\tau$ symmetry. 
On the other hand, the assumption of a 
(maximally allowed) four zero texture in the Yukawa coupling matrix $Y_\nu$ in the same basis leads to a 
highly constrained
and predictive theoretical scheme. We show that the requirement of an exact $\mu\tau$ symmetry, 
coupled with observational constraints,
reduces the {\it seventy two} allowed textures in such a $Y_\nu$ to {\it only four} 
corresponding to just two different forms of the light neutrino mass matrix $m_\nu$. 
The effect of each of these on measurable quantities can be described, 
apart from an overall factor of the neutrino mass scale, 
in terms of two real parameters and a phase angle all of which are within very 
constrained ranges. The additional 
input of a tribimaximal mixing reduces these three parameters to {\bf only one} with a very nearly 
fixed value. 
Implications for both flavored 
and unflavored leptogenesis as well as radiative lepton flavor violating decays are discussed.  
We also investigate the stability of these conclusions 
under small deviations 
due to renormalization group running from a high scale where 
the four zero texture as well as 
$\mu\tau$ symmetry or the tribimaximal mixing pattern are imposed. 
\end{abstract}
PACS number(s): 14.60.Pq, 11.30.Hv, 98.80.Cq
\section{Introduction}
A lot is now known \cite{camelliari} about the masses and mixing angles of the three 
light neutrinos, based on the solid foundation 
of accumulated experimental evidence, while the remaining gaps are expected to be filled in the 
foreseeable  future. 
Thus the task of pinning down the form of their Yukawa coupling matrix $Y_\nu$ in flavor space, 
assuming the existence of three heavy right chiral neutrinos, is very much at hand.  
The general structure of $Y_\nu$ is, however, intractable at the moment. One needs concrete theoretical ideas 
to simplify it and then test such simplified forms by comparing with extant data. Our present work is in 
such a spirit. 
\vskip 0.1in
\noindent
 We try in this paper to bring together three theoretical ideas : (1) allowed four 
zero neutrino Yukawa textures
\cite{bc}-\cite{ctb}, (2) $\mu\tau$ 
symmetry 
\cite{Fuku}-\cite{Ghosal} 
and 
(3) a tribimaximal mixing pattern\footnote{ Such a pattern could be due to a 
flavor symmetry in the Lagrangian such as  $A_4$ \cite{alta1}, 
$S_3$ \cite{alta2}.}
\cite{Har}-\cite{Harri}, 
which actually subsumes the results of (2). Within the type-I seesaw 
framework \cite{Min}-\cite{moha} 
and in the weak 
basis where the charged leptons $l_\alpha$ ($\alpha$ = 1,2,3) and the 
heavy right chiral neutrinos $N_i$ (i=1,2,3)
have real and diagonal respective masses $m_\alpha$ 
and $M_i$, we explore the mutual compatibility between 
(1) and (2) and further 
between (1) and (3). 
A drastic reduction of the allowed textures and parameters under (1) ensues.
\vskip 0.1in
\noindent
Let us start with (1). Assuming the absence{\footnote 
{Allowing one massless neutrino, five zeroes 
are allowed in $Y_\nu$ 
\cite{Goswami:2009bd}-\cite{Dighe:2009xj}.}} 
of any strictly massless neutrino as well as that of any 
unnatural cancellation, 
the utilization of the observed lack of complete decoupling of any neutrino flavor 
from the two others led to the demonstration \cite{bc} that four is the maximum number of zeroes 
allowed in $Y_\nu$. All allowed four zero textures, seventy two configurations in total, were completely 
classified in \cite{bc} into two categories : (A) fifty four textures with two (element by 
element) orthogonal rows $i$ and $j$ say; (B) eighteen textures with 
nonorthogonal rows and one row having two zeroes 
with the other two rows ($k$ and $l$, say) having one zero each. Let us write the complex symmetric light 
neutrino Majorana mass matrix in our basis as
\be
m_\nu = -Y_\nu \,\,{\rm diag.}(M_1^{-1}, M_2^{-1}, M_3^{-1})\,\,Y_\nu^T v^2,
\ee
{\it{v}} being the relevant Higgs VEV. Now, for all textures  of category (A), one has the condition 
\cite{bc}
\be
{(m_\nu)}_{ij} = 0  : \rm{category\,(A)},
\ee
while, for those of category (B), the condition is \cite{bc}
\be
 {\rm{det\,\, cofactor}} [{(m_\nu)}_{kl}] = 0   :  \rm{category\,(B)}.
\ee
\vskip 0.1in
\noindent
One very important and interesting feature of all these allowed 
four zero textures is that they enable 
\cite{bc} the complete reconstruction of the neutrino Dirac mass matrix 
$m_D = vY_\nu$ in terms of the physical masses of the light 
neutrinos as well as $M_{1,2,3}$ and the elements of the unitary PMNS mixing matrix including the Majorana 
phase matrix factor. This means \cite{bc} that the high scale CP violation required for 
leptogenesis gets specified exclusively 
\cite{Frampton:2002yf}-\cite{Babu:2008kp}  
in terms of the CP violation pertaining to laboratory 
energy neutrinos. Another striking feature of these textures 
is the following. 
Conditions (1.2) and (1.3) on the corresponding neutrino mass matrix $m_\nu$ are 
invariant \cite{ctb} under renormalization group running at the one loop level, though 
texture zeroes in general are not. Thus if these conditions are the 
consequences of some symmetry operative at a high scale, 
they would be approximately 
valid {\it even\, at\, laboratory\, energies} 
where neutrino oscillation experiments are performed.
\vskip 0.1in
\noindent
We next come to (2), i.e, $\mu\tau$ symmetry 
\cite{Fuku}-\cite{Ghosal}. For the purpose of implementing it, 
we find it convenient to choose the following representation of the PMNS mixing matrix
\be
U= \pmatrix{c_{12}c_{13}&
                      s_{12}c_{13}&
                      -s_{13}e^{-i\delta}\cr
-s_{12}c_{23}-c_{12}s_{23}s_{13}e^{i\delta}&c_{12}c_{23}-s_{12}s_{23}s_{13}e^{i\delta}&
-s_{23}c_{13}\cr
-s_{12}s_{23} +c_{12}c_{23}s_{13}e^{i\delta}&c_{12}s_{23} +s_{12}c_{23}s_{13}e^{i\delta}&
c_{23}c_{13}\cr
}
\pmatrix{1&0&0\cr
         0&e^{i\alpha_M}&0\cr
         0&0&e^{i(\beta_M+\delta)}},
\ee
with $c_{ij} = cos\,\theta_{ij}$, $s_{ij} = sin\,\theta_{ij}$, 
$\delta$ being the Dirac phase and $\alpha_M$, $\beta_M$ being the Majorana phases.
It is important to note that, with three real neutrino mass 
eigenvalues $m_{1,2,3}$, one has
\be
                   m_\nu = U\,\, {\rm diag.}(m_1, m_2, m_3)\,\, U^T.
\ee
We can now define $\mu\tau$ symmetry to be the invariance of all 
couplings and masses in the pure 
neutrino part{\footnote{This symmetry is, of course, badly broken in the 
charged lepton sector.}} of the 
Lagrangian under the interchange of the flavor indices 2 and 3. 
As a result, 
\be
{(Y_\nu)}_{12} = {(Y_\nu)}_{13}, 
\ee
\be
{(Y_\nu)}_{21}={(Y_\nu)}_{31}, 
\ee
\be
{(Y_\nu)}_{23} = {(Y_\nu)}_{32},
\ee
\be
{(Y_\nu)}_{22}  = {(Y_\nu)}_{33} 
\ee 
and 
\be
M_2 = M_3. 
\ee
Using eq.(1.1), one then obtains 
\be
{(m_\nu)}_{12} = {(m_\nu)}_{13},               
\ee
\be
{(m_\nu)}_{22} = {(m_\nu)}_{33}.               
\ee
We shall take eqs.(1.11, 1.12) as the statement of a custodial $\mu\tau$ symmetry of the light neutrino mass 
matrix $m_\nu$.
An automatic consequence of these two equations is the 
fixing of the two mixing angles involving the third flavor at 
$\theta_{23}= \pi/4$, $\theta_{13}=0$. Discarding unnatural cancellations, {\it sixty eight} of the 
{\it seventy two} allowed four zero textures in $Y_\nu$ are 
found to be incompatible with eqs.(1.11, 1.12) plus observational constraints. In 
particular, fifty two textures of category (A) and sixteen textures of (B) 
category are excluded. The two surviving textures of category A both lead to 
the same light neutrino mass matrix with $(m_\nu)_{23} = 0$. On the other hand, each of the two surviving 
category (B) textures turns out to have two zeroes in the first row and one each in the other rows  
and they also lead to the same light neutrino mass matrix. 
For each surviving texture, $m_\nu$ can be described, 
apart from an overall neutrino mass scale, by two real parameters 
and one phase angle, though their definitions are different for category A and category B. We call them $k_1$, $k_2$ and $\alpha$ for the former and $l_1$, $l_2$ and $\beta$ for the latter. Their allowed ranges
are found to be severely constrained by the neutrino oscillation 
data. 
\vskip 0.1in
\noindent
We then turn to the 
tribimaximal mixing (TBM) pattern 
\cite{Har}-\cite{Harri} which implies $\theta_{13}$ = 0, $\theta_{23} = \pi/4$ 
 {\it and}\, 
$\theta_{12} = {\sin}^{-1}/\sqrt{3}$. The effect of $\mu\tau$ 
symmetry is thus subsumed here, but 
there is an additional constraint on $\theta_{12}$. Hence all configurations of $m_\nu$ leading to TBM 
have not only to obey eqs.(1.11-1.12) but also the extra requirement 
\be
{(m_\nu)}_{11} + {(m_\nu)}_{13} = {(m_\nu)}_{22} + {(m_\nu)}_{23}.
\ee
The four textures of $Y_\nu$, allowed by  $\mu \tau$ symmetry, 
survive the imposition of eq.(1.13), 
but two relations between $k_1$, $k_2$ and 
$\alpha$ for category A and two between $l_1$, $l_2$ and $\beta$ for category B emerge. 
Consequently, one independent real parameter $k_2$ for the former and one $l_1$ for the latter suffice 
to describe those textures after factoring out the overall mass scale. 
The allowed domains of $k_2$ and $l_1$  
are again found to be highly restricted.    
\vskip 0.1in
\noindent
A general nondiagonal Majorana mass matrix $m_\nu$ in flavor space implies lepton flavor 
violation as well as the nonconservation of 
lepton number. It is therefore interesting 
and important to discuss the implications of the 
above forms of $m_\nu$ for\footnote{Nonradiative lepton flavor violating 
processes, such as $\mu e$ conversion in nuclei and triple charged 
leptonic decays of the $\tau$ and the $\mu$, are not 
considered here since current experimental limits on those 
yield considerably weaker constraints than radiative 
lepton flavor violating decays.} radiative lepton flavor 
violating decays $(\tau\rightarrow \mu\gamma, 
\,\, \tau\rightarrow e\gamma, \,\,\mu\rightarrow e\gamma)$ and for realistic 
leptogenesis of both flavor independent and flavor dependent varieties. 
The former are yet-to-be-observed processes \cite{add1} for which the experimental sensitivity 
is rapidly approaching theoretical expectations; the latter is a desirable theoretical 
goal \cite{add2} of any (high scale) seesaw-based model of 
light neutrino masses and mixing angles. 
In the mSUGRA version \cite{bookproy} of a supersymmetric scenario, 
the branching ratios for the three radiative lepton flavor 
violating decays in question have rather simple flavor structures 
that are bilinear in $Y_\nu$ or $m_D$. 
We are thus able to make some specific predictions for our allowed textures, 
namely, the vanishing of BR$(\tau\rightarrow \mu\gamma)$ for category A and the value of the ratio 
BR$(\tau\rightarrow e\gamma)$/BR$(\mu\rightarrow e\gamma)$ being $\simeq $ 0.178 for 
both categories. 
Concerning leptogenesis, 
the term contributing only to flavor dependent lepton asymmetries 
vanishes  for all flavor combinations in both categories. 
Regarding the term, which contributes to the flavor summed lepton asymmetry, 
only the electron asymmetry gets generated in category A whereas the same always 
vanishes in category B. One can also make more definitive statements 
on specific flavor combinations of the latter term as well as on the 
effective mass for the washout of a particular flavor asymmetry. 
\vskip 0.1in
\noindent
One issue with $\mu\tau$ symmetry and TBM is that the former 
fixes $\t13$ and $\ta23$ at 0 and $\pi/4$ respectively, 
while the latter further fixes $\ts12$ at $\sin^{-1}\frac{1}{\sqrt{3}}$   
$\simeq$ ${35.26}^\circ$. Though these numbers lie within presently allowed 
$3\sigma$ ranges of those  mixing angles, 
the true values of the latter may eventually turn out to be different. 
There are, in fact, hints already that such may be the case. Current best fit $1\sigma$ ranges
for those angles, derived from global analyses of all neutrino oscillation data, are 
\cite{add3} $\ts12 = 34.5^\circ\pm 1.4^\circ$, $\ta23 = {43.1^\circ}^{+4.4^\circ}_{-3.5^\circ}$ and 
$\t13 = 8^\circ\pm 2^\circ$. While it is premature to take these ranges too seriously, it is nonetheless 
interesting to consider deviations within a definitive theoretical framework by taking them 
to originate dynamically from radiative effects. We impose $\mu\tau$ symmetry or TBM on elements 
of the light neutrino mass matrix $m_\nu$ at a high scale of the order of the lowest heavy right 
chiral neutrino mass, i.e. at $\Lambda\sim\,\,$min$(M_1, M_2, M_3)$\, 
$\sim$\,\, ${10}^{12}$ GeV. 
We further assume the validity of the Minimal Supersymmetric Standard Model (MSSM) \cite{bookproy} 
between this scale and the laboratory energy scale $\lambda\sim {10}^{3}$ GeV. 
The elements of $m_\nu$ are then evolved from 
$\Lambda$ to $\lambda$ by one loop renormalization group running. 
Small deviations from the consequences of $\mu\tau$ symmetry or TBM, 
proportional to the square of the heaviest charged lepton mass 
divided by the Higgs VEV squared, are found to be generated. 
These lead to small but distinct 
extensions of the allowed 
values of $k_{1,2}$ in category A and $l_{1,2}$ in category B. 
Constrained deviations in 
the mixing angles also emerge. 
\vskip 0.1in
\noindent
The rest of the paper is organized as follows. 
Section 2 contains a discussion of the allowed four zero 
textures and their parameterization as a consequence of 
$\mu\tau$ symmetry and TBM. 
Radiative lepton flavor violating decays and leptogenesis 
are taken up for those textures in Section 3. In Section 4, 
radiatively induced small deviations in $m_\nu$ and 
their effects are discussed. 
The final Section 5 contains a summary of our results and the conclusions 
derived therefrom. 
The Appendix contains analytical expressions for the 
experimentally measured quantities 
utilized by us both without and with one loop RG evolution.
\section{Allowed four zero textures}
{\bf Category A}
\vskip 0.1in
\noindent
It is straightforward to see that only two of the fifty two four zero textures of category (A) are consistent with 
 $\mu\tau$ symmetry, as implemented through eqs. (1.11, 1.2). 
The rest develop additional zeroes which are incompatible 
with known observational constraints and the assumption of no massless neutrino. 
The two allowed textures for 
the Dirac mass matrix $m_D = Y_\nu v$ can be given in terms of three complex parameters 
$a_1$, $a_2$, $b_1$ as 
\be
m_D^{(1)} = \pmatrix{a_1&a_2&a_2\cr
                     0&0&b_1\cr
                     0&b_1&0},
\ee
\be
m_D^{(2)} = \pmatrix{a_1&a_2&a_2\cr
                     0&b_1&0\cr
                     0&0&b_1}.
\ee 
The corresponding light neutrino mass matrices are 
identical and can be written as 
\be 
m_\nu^{(A)}= -\pmatrix{a_1^2/M_1 + 2a_2^2/M_2 & a_2b_1/M_2 & a_2b_1/M_2\cr
                       a_2b_1/M_2 & b_1^2/M_2 &0\cr
                       a_2b_1/M_2 & 0 & b_1^2/M_2}.
\ee
\noindent 
Let us now define 
$m \equiv -\frac{b_1^2}{M_2}$, $k_1e^{i(\alpha+\alpha^\prime)} \equiv 
\frac{a_1}{b_1}\frac{\sqrt M_2}{\sqrt M_1}$, 
$k_2e^{i\alpha^\prime} \equiv \frac{a_2}{b_1}$ and 
further absorb the phase $\alpha^\prime$ in the first family 
neutrino field $\nu_e$. 
The latter is equivalent to rotating the mass matrix of eq.(2.3) 
by the phase matrix {\rm diag.}$(e^{-i\alpha^\prime}, 1,1)$. 
This operation changes eq.(2.3) to 
\be
m_\nu^{(A)} = m \pmatrix{k_1^2e^{2i\alpha}+2k_2^2&k_2&k_2\cr
                                    k_2&1&0\cr
                                    k_2&0&1}.
\label{mnuA} 
\ee
Apart from the overall mass scale factor $m$, 
the light neutrino mass matrix now has two real parameters $k_1$, $k_2$ and the phase angle 
$\alpha$. 
\par
The ratio $R = \Delta m_{21}^2/\Delta m_{32}^2$
and the solar/reactor mixing angle $\theta_{12}$ are now given by
\be
R = 2{(X_1^2+X_2^2)}^{1/2}{[X_3-{(X_1^2+X_2^2)}^{1/2}]}^{-1},
\ee
\be
\tan2\theta_{12} = \frac{X_1}{X_2}
\ee
\noindent
with 
\be 
X_1 = 2\sqrt{2}k_2{[{(1+2k_2^2)}^2 + k_1^4 + 2k_1^2(1+2k_2^2)\cos2\alpha]}^{1/2},
\ee
\be
X_2 = 1-k_1^4-4k_2^4-4k_1^2k_2^2\cos2\alpha,
\ee
\be
X_3 = 1-4k_2^4-k_1^4-4k_1^2k_2^2\cos2\alpha - 4k_2^2. 
\ee
The observables of eqs.(2.5) and (2.6) can be compared with the 
available data. We see right away that the expression for $R$ is 
{\it incompatible with a normal mass ordering 
($\Delta m_{32}^2 > 0$) 
and can only accommodate an inverted one ($\Delta m_{32}^2 < 0$)}. 
This is 
consistent with the conclusion of 
Merle and Rodejohann \cite{Merle:2006du} who had shown that the 
condition 
${(m_\nu)}_{23} = {(m_\nu)}_{32} = 0$ is 
compatible only with an inverted mass ordering.   
The allowed ranges are 
given respectively{\footnote{We are using the range of R extracted 
\cite{Maltoni:2008ka} by assuming an inverted mass-ordering.}}
by $R = -3.476\times10^{-2}\,\,\rm{eV}^2$ to 
\begin{figure}[htb]
\hskip 1.20cm
\psfig{figure=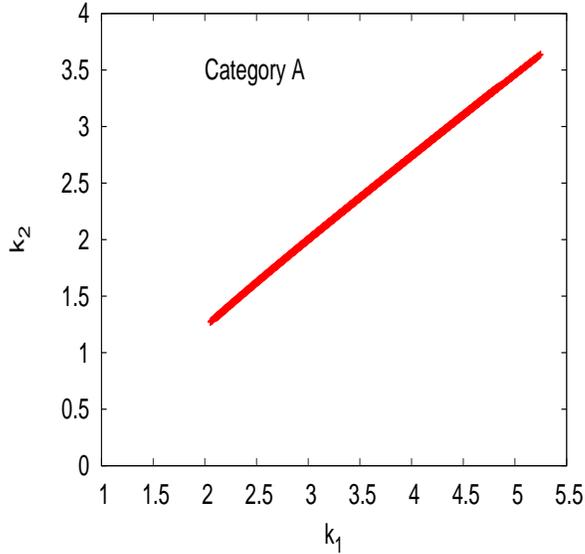,width=8.0cm,height=8.0cm,angle=270}
\caption{\small
Variation of $k_1$ and $k_2$ in category A with $\mu\tau$ symmetry over the $3\sigma$ allowed ranges of 
$R$ and $\theta_{12}$.  
}
\end{figure}
$-2.972\times 10^{-2}\,\,\rm{eV}^2$ at the $1\sigma$ level and 
$-4.129\times10^{-2}\,\,\rm{eV}^2$ to
$-2.534\times 10^{-2}\,\,\rm{eV}^2$ at the $3\sigma$ level and
by
$\tan2\theta_{12}$ = 3.045 - 2.278 at  1$\sigma$ and  
4.899 - 1.828 at 3$\sigma$. 
The angle 
$\alpha$ is immediately found to be 
correspondingly restricted to be between $89^\circ$ and 
$90^\circ$. 
We find that 
{\it there is no acceptable solution for the $1\sigma$-allowed range of $R$}. 
For the 
$3\sigma$-allowed range of the latter, 
a very narrow strip is allowed in the $k_1$-$k_2$ plane 
for the allowed domain of 
$\alpha$, as shown in Fig.1 with $ 2.0 < k_1 < 5.3$ and $1.2 < k_2 <3.7$. 
Thus $m_\nu^{(A)}$ may quite possibly be excluded by 
further improvements of error 
in the data on $R$ and $\tan 2\theta_{12}$.
\par
On further assuming 
tribimaximal neutrino mixing, i.e, eq.(1.13), one obtains the relation 
\be
k_1^2e^{2i\alpha} + 2k_2^2 + k_2 = 1.
\ee
Given eq.(2.10), $\alpha$ is now fixed{\footnote {The solution $\alpha$ = 0 is incompatible with the 
allowed range of $R$ and the 
reality of $k_{1,2}$.}}  
to be $\pi/2$ and the two real parameters $k_{1,2}$ are 
therefore reduced to one, 
which we take to be $k_2$ fixing $k_1$ at 
\be
k_1 = {(2k_2^2 + k_2 - 1)}^{1/2}.
\ee
\noindent
Now that $\tan2\theta_{12}$ is fixed at $2\sqrt{2}$, the ratio $R$ is 
given by
\be
R =\frac{3(k_2 - 2)}{k_2+2}.
\ee
The range of $k_2$ restricted by the $3\sigma$ allowed domain of 
$R$ is now $1.95\leq k_2 \leq 1.97$ so that its value is fixed to the first decimal place.
\vskip 0.1in
\noindent
{\bf Category B}
\vskip 0.1in
\noindent
Again, in this case, only two of the original eighteen 
textures are allowed by $\mu\tau$ symmetry.   
These may be written in terms of three complex parameters $a_1$, $b_1$, $b_2$ as  
\be
m_D^{(3)} = \pmatrix{a_1&0&0\cr
                     b_1&b_2&0\cr
                     b_1&0&b_2},
\ee
\be
m_D^{(4)} = \pmatrix{a_1&0&0\cr
                     b_1&0&b_2\cr
                     b_1&b_2&0},
\ee
with the corresponding light neutrino mass matrices both being 
\be
m_\nu^{(B)} = -\pmatrix{a_1^2/M_1 & a_1b_1/M_1 & a_1b_1/M_1\cr
                        a_1b_1/M_1 & b_1^2/M_1+b_2^2/M_2 &b_1^2/M_1\cr
                        a_1b_1/M_1 & b_1^2/M_1 & b_1^2/M_1+b_2^2/M_2}.
\ee
\noindent
Now, we choose to define 
$m = -\frac{b_2^2}{M_2}$, $l_1e^{i\beta^\prime} = 
\frac{a_1}{b_2}\frac{\sqrt M_2}{\sqrt M_1}$,
$l_2e^{i\beta} = \frac{b_1}{b_2}\frac{\sqrt M_2}{\sqrt M_1}$ 
and absorb the phase $\beta^\prime$ in $\nu_e$. We are then left with 
\be
m_\nu^{(B)} = m \pmatrix{l_1^2&l_1l_2e^{i\beta}&l_1l_2e^{i\beta}\cr
                                    l_1l_2e^{i\beta}&l_2^2e^{2i\beta}+1&l_2^2e^{2i\beta}\cr
                                    l_1l_2e^{i\beta}&l_2^2e^{2i\beta}&l_2^2e^{2i\beta}+1}. 
\label{mnuB} 
\ee
The measurable quantities $R$ and $\tan2\theta_{12}$ are still given by eqs.(2.5) and (2.6), but now 
the functions $X_{1,2,3}$ are given in terms of the parameters $(l_1,\, l_2\,, \beta)$ as 
\be
X_1 = 2\sqrt{2}l_1l_2{[{(l_1^2+2l_2^2)}^2 + 1+ 2(l_1^2+2l_2^2)\cos2\beta]}^{1/2},
\ee
\be
X_2 = 1+4l_2^2\cos2\beta+4l_2^4-l_1^4,
\ee
\be
X_3 = 1-{(l_1^2+2l_2^2)}^2 - 4l_2^2\cos2\beta.
\ee
In this case we see that the 
expression for $R$ {\it admits only a normal mass ordering and disallows 
an inverted one}.  
A comparison with data fixes $\beta$ in the ranges $89^\circ$ to $90^\circ$ and 
$87^\circ$ to $90^\circ$ respectively for the values 
of{\footnote{We are using the range of R 
extracted \cite{Maltoni:2008ka} by assuming a normal mass-ordering.}}
$R = 
3.329\times10^{-2}\,\,\rm{eV}^2$ to
$2.858\times 10^{-2}\,\,\rm{eV}^2$ at the $1\sigma$ level and
$3.915\times10^{-2}\,\,\rm{eV}^2$ to
$2.455\times 10^{-2}\,\,\rm{eV}^2$ at the  
$3\sigma$ level with the allowed values 
of $\tan2\theta_{12}$ as previously mentioned. 
\begin{center}
\begin{figure}[htb]
\hskip 1.20cm
\psfig{figure=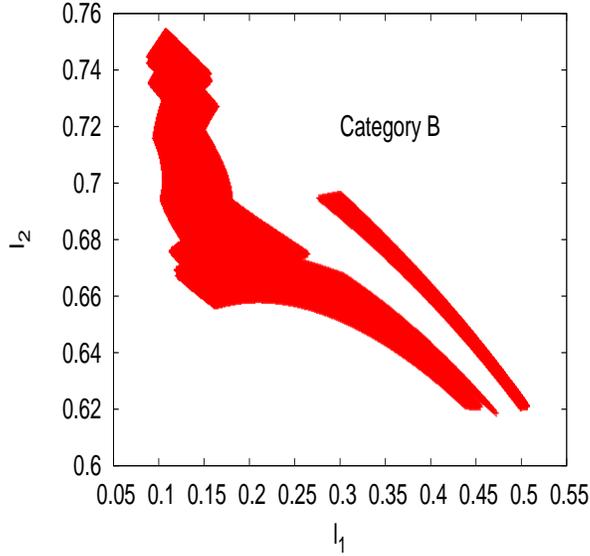,width=8.0cm,height=8.0cm,angle=270}
\caption{\small
Variation of $l_1$ and $l_2$ in category B with $\mu\tau$ symmetry over the $3\sigma$ allowed ranges of 
$R$ and $\theta_{12}$.}
\label{f:ecal}
\end{figure}
\end{center}
The corresponding allowed values of $l_{1,2}$ are 
shown in Fig.2 for the $3\sigma$-allowed range.
Unlike category A, a substantial region of the parameter space, consisting of two branches, is allowed here. 
\vskip 0.1in
\noindent
The imposition of the tribimaximal mixing condition of eq.(1.13) now leads to 
\be 
l_1^2 + l_1l_2e^{i\beta} = 2l_2^2e^{2i\beta} + 1
\ee
which fixes $\beta$ by{\footnote{
The solution $\beta= 0$ is not compatible with real $l_{1,2}$ and the allowed range 
of $R$.}}  
\be
\cos\beta = \frac{l_1}{4l_2}.
\ee
Moreover, $l_{1,2}$ can now be reduced to a single real parameter $l_1$ with $l_2$ given by  
\be
l_2 = {\frac{1}{2}(1-l_1^2)}^{1/2}.
\ee
Again, $\tan2\theta_{12}$ being $2\sqrt{2}$, R is given by 
\be
R = \frac{3l_1^2}{2-4l_1^2}.
\ee
In consequence, the allowed $1\sigma$ and $3\sigma$ ranges of $l_1$ get 
restricted to $0.12\leq l_1 \leq 0.13$ 
and $0.11\leq l_1 \leq 0.15$ respectively. Once again, the value of this surviving one parameter is fixed to the first decimal place.
\section{Radiative lepton flavor violation and leptogenesis}
Radiative lepton flavor violating decays $l_\alpha\rightarrow l_\beta\gamma$ 
(flavor indices $\alpha,\beta$ spanning $1=e, 2=\mu, 3=\tau$ with 
the constraint 
$\alpha>\beta$) together with the required generation of a 
lepton asymmetry at a high scale, 
provide powerful tools to check and test \cite{Davidson:2001zk}-\cite{Ibanez:2009du}
any proposed seesaw-based scheme 
of neutrino mixing and masses. There already exist lower bounds on the partial 
lifetimes of the former processes; moreover, forthcoming experiments with higher sensitivity 
 will hope 
to observe some of the decay channels. Coming to leptogenesis as a route to baryogenesis, a 
fair amount of theoretical 
understanding exists for high scale leptogenesis - both of the flavored 
and unflavored varieties. In this section, we explore the implications 
of the allowed four zero texture configurations, with tribimaximal mixing or at 
least $\mu\tau$ symmetry, for these two types of phenomena.
\vskip 0.1in
\noindent
We note first the one-loop expression \cite{Borzumati:1986qx} for 
BR($l_\alpha\rightarrow l_\beta\gamma$) 
which is valid in mSUGRA scenarios with universal boundary 
conditions on the masses of scalar particles at a 
high scale $M_X$ : 
\be 
{\rm{BR}}(l_\alpha\rightarrow l_\beta\gamma)= 
{\rm const.}\, {\rm BR}(l_\alpha\rightarrow l_\beta\nu{\bar\nu})
|(m_DLm_D^\dagger)_{\alpha\beta}| 
\ee
with 
\be
L_{kl} = \ln\frac{M_X}{M_k}\delta_{kl},
\ee
$M_k$ being the mass of the $k$th. heavy right chiral neutrino. The matrix $L$ takes care of the 
RG running from $M_X$ to $M_k$. We can now discuss what happens with our four allowed configurations for 
$m_D$.
\vskip 0.1in
\noindent
{\bf Category A}
\vskip 0.1in
\noindent
For both the allowed textures $m_D^{(1)}$ and $m_D^{(2)}$, we have
\be
{(m_\nu)}_{23} = -{(m_D M_{R}^{-1}m_D^T)}_{23} = 0
\ee
in a basis in which $M_R$ is diagonal. Since $L$ is a diagonal matrix, it follows that 
\be
{(m_D L m_D^\dagger)}_{23} =0.
\ee
Consequently,
\be
{\rm{BR}}(\tau\r \mu\gamma) = 0.
\ee
and any observation of the $\tau\r \mu\gamma$ process will rule out these configurations. It has moreover
been shown \cite{Merle:2006du}
 from the twin requirements of two nonzero neutrino masses and mixing angles 
that in such a case ${(m_\nu)}_{12} \neq 0 \neq {(m_\nu)}_{13}$.  As a result,  
${(m_D L m_D^\dagger)}_{12}$ and ${(m_D L m_D^\dagger)}_{13}$  are also both nonzero, leading to 
nonvanishing rates for the decays 
$\mu\r e\gamma$ and  $\tau\r e\gamma$ respectively.  There is moreover a relation 
between them.  On account of $\mu\tau$ symmetry, $M_2 = M_3$ and  ${(m_D L m_D^\dagger)}_{12}$ = 
 ${(m_D L m_D^\dagger)}_{13}$,  so that we have
\be 
\frac{{\rm BR}(\tau\r e \gamma)}{{\rm BR}(\mu\r e \gamma)}\simeq 
\frac{\rm{BR}(\tau\r e\nu_\tau\bar\nu_e)}{\rm{BR}(\mu\r e\nu_\mu\bar\nu_\mu)} \simeq 0.178.
\ee
\vskip 0.1in
\noindent
{\bf Category B}
\vskip 0.1in
\noindent       
For both the allowed textures  
$m_D^{(3)}$ and  $m_D^{(4)}$, the matrix $m_D L m_D^\dagger$ is 
identical with all elements nonvanishing. Thus, all the three radiative modes 
$\mu\r e \gamma$, $\tau\r \mu \gamma$, $\tau\r e\gamma$ are possible. However, $\mu\tau$ symmetry 
has the same consequence as in category A, i.e eqn.(3.6) holds here too. 
\vskip 0.1in
\noindent
We next turn to leptogenesis at the scale $\sim$ min($M_1$, $M_2$, $M_3$) 
which for simplicity we take 
to be $M_1$. Most pertinent for this are the lepton asymmetries generated by 
the decay of a heavy right 
chiral neutrino $N_i$ into a lepton of flavor $\alpha$ 
(= e, $\mu$, $\tau$) and a Higgs $\phi$: 
\bea
\e_i^\alpha &=& \frac{\Gamma(N_i\r \phi\bar{l}_{\alpha})-\Gamma(N_i\r\phi^\dagger l_\alpha)}
{\sum_{\beta} [\Gamma(N_i\r\phi\bar{l_\beta}) + \Gamma(N_i\r \phi^\dagger l_\beta)]}
\nonumber\\
&\simeq& 
\frac{g^2}{16\pi M_W^2}\frac{1}{{(m_D^\dagger m_D)}_{ii}}\sum_{j\neq i}\left[{\cal I}^\alpha_{ij} 
f\left(\frac{M_j^2}{M_i^2}\right) + {\cal J}^\alpha_{ij}{\left(1- \frac{M_j^2}{M_i^2}
\right)}^{-1}
\right] 
\eea  
where we have neglected $O(M_W^2/M_i^2)$ terms. Here 
\be
{\cal I}^\alpha_{ij} = {\rm{Im}}{(m_D^\dagger)}_{i\alpha}{(m_D)}_{\alpha j}{(m_D^\dagger m_D)}_{ij} 
= -{\cal I}^\alpha_{ji}
,$$
$$
{\cal J} ^\alpha_{ij} = {\rm{Im}} 
{(m_D^\dagger)}_{i\alpha}{(m_D)}_{\alpha j}{(m_D^\dagger m_D)}_{ji} = 
-{\cal J}^\alpha_{ji}.
\ee
The function $f(x)$ has the form 
\be
f(x) = \sqrt{x}\left[\frac{2}{1-x} - \ln\frac{1+x}{x}\right]
\ee 
in the MSSM. For $M_1<< M_{2,3}$, $f(M_{2,3}^2/M_1^2)\simeq -3 M_1/M_{2,3}$
in which case the  ${\cal J}^\alpha_{ij}$ term in $\e_i^\alpha$ gets suppressed by $M_1/M_{2,3}$. 
Another interesting quantity is the effective mass for the washout 
of a flavor asymmetry. This is given by 
\cite{Covi:1996wh}- \cite{Nardi:2006fx}  
\be
{\tilde{m}}_1^\alpha = |{(m_D)}_{\alpha1}|^2/M_1
\ee
and controls the magnitude of the final baryon asymmetry $Y_B$ in the way shown in Ref.
  \cite{Covi:1996wh} - \cite{Nardi:2006fx}.
Summing over all lepton flavors $\alpha$, the ${\cal J}^{ij}_\alpha$ term drops out since 
$\sum_\alpha {\cal J}^\alpha_{ij} = 0$. 
Utilizing the result that ${\cal I}_{ij} = \sum_\alpha {\cal I}_{ij}^\alpha$
$= {\rm{Im}}{[{(m_D^\dagger m_D)}_{ij}]}^2$, we have 
\be
\e_i = \sum_\alpha \e_i^\alpha = \frac{g^2}{16\pi M_W^2}\frac{1}{{(m_D^\dagger m_D)}_{ii}} 
\sum_{j\neq i}{[{(m_D^\dagger m_D)}_{ij}]}^2 f\left(M_j^2/M_i^2\right).
\ee 
Though the above expressions are valid in the MSSM, their flavor structure is just that of 
the Standard Model.
\vskip 0.1in
\noindent
Selecting the $\mu\tau$ symmetric four zero texture configurations of $m_D$, we find that 
${\cal J}_{ij}^\alpha$ vanishes in every case for all $\alpha$, $i$, $j$. Thus we need 
not consider the second term in eqn.(3.7) at all. Regarding ${\cal I}_{ij}^\alpha$, both allowed 
textures in category A yield the same result, namely, ${\cal I}_{12}^e = {\cal I}^e_{13}\neq 
0$ while the other combinations vanish. Therefore, only the electron asymmetry gets 
generated in this case. In category B, $m_D^{(3)}$ leads to 
nonzero and equal ${\cal I}^\mu_{12}$, 
${\cal I}^\tau_{13}$, with all other ${\cal I}_{ij}^\alpha$ vanishing, while $m_D^{(4)}$  
yields nonvanishing and equal ${\cal I}^\mu_{13}$,  ${\cal I}^\tau_{12}$, the rest of 
${\cal I}_{ij}^\alpha$ being zero. Turning to the effective washout mass,
only the electron one, namely $\widetilde{m_1}^e$, is nonvanishing for both textures of 
category A. For those of category B, all the washout masses $\widetilde{m_1}^e$, 
$\widetilde{m_1}^\mu$, $\widetilde{m_1}^\tau$ are nonzero with 
$\widetilde{m_1}^\mu$ = $\widetilde{m_1}^\tau$. We provide a table 
containing the 
relevant information on leptogenesis parameters for each of our allowed four texture 
zero configurations.
\begin{center}
\begin{table}[htb]
\begin{tabular}{|c|c|c|c|c|c|}
\hline
configuration & ${\cal I}_{ij}^\alpha$& ${\cal J}_{ij}^\alpha$&$\widetilde{m_1}^e$
&$\widetilde{m_1}^\mu$&
$\widetilde{m_1}^\tau$\\
\hline
$m_D^{(1)}$ & ${\cal I}^e_{12}={\cal I}^e_{13}\neq 0$, \rm{rest zero}&0&nonzero&0&0\\
\hline
$m_D^{(2)}$ & --do-- & 0&nonzero&0&0\\
\hline
$m_D^{(3)}$ & ${\cal I}^\mu_{12}={\cal I}^\tau_{13}\neq 0$, \rm{rest zero}&0&nonzero&nonzero&equals 
$\widetilde{m_1}^\mu$\\
\hline
$m_D^{(4)}$ & ${\cal I}^\mu_{13}={\cal I}^\tau_{12}\neq 0$, \rm{rest zero}& 0 &nonzero &nonzero &equals 
$\widetilde{m_1}^\mu$\\
\hline
\end{tabular}
\caption{Leptogenesis Table}
\end{table}
\end{center}
\section{Radiatively induced deviations}
We mentioned 
in the previous section that the results 
$\t13= 0$ and $\ta23 = \pi/4$ follow from a 
custodial $\mu\tau$ symmetry in $m_\nu$. A  
breaking of this symmetry would in general result in a  nonzero 
value of $\t13$ as well as a departure of $\ta23$ from $\pi/4$. The goals of 
many ongoing and planned experiments are to measure their actual values \cite{plan}. 
Another interesting consequence of a nonzero $\t13$ would be the presence of a CKM-type 
of CP violation in the lepton sector.
Our previous expressions for $R$ 
and $\tan2\ts12$ will be modified if $\mu\tau$ symmetry is indeed broken.

\noindent
In this section we invoke the dynamical origin of such a 
symmetry breaking due to the Renormalization Group (RG) 
 evolution of the elements of the 
neutrino mass matrix. 
Our basic idea is to posit that $\mu\tau$ symmetry 
(or more restrictively whichever symmetry, say $A_4$ or $S_3$ is responsible 
for TBM) is valid at a high energy scale $\Lambda\sim 10^{12}$ GeV which characterizes the heavy 
right chiral neutrinos $N_i$. We then consider the 
radiative breaking of such a 
symmetry through charged lepton mass 
terms, 
induced at the one loop level, 
as one evolves by RG running to the lower energy scale $\lambda\sim 10^3$ GeV. 
The specific theory in which we choose to do 
this is the Minimal Supersymmetric Standard Model 
(MSSM) \cite{bookproy} with an intrasupermultiplet mass splitting, 
caused by explicit 
supersymmetry breaking, being $O(\lambda)$.
Following the methodology described in Ref. \cite{dgr1}-\cite{dgr2}, 
we consider the neutrino mass matrices $m_nu$ given in eqns.(2.4) and (2.16) at the 
high scale $\Lambda$. Their evolved form at the 
low scale 
$\lambda$ is then given approximately by{\footnote{In terms of $Y_\nu$ with which we started, 
$Y_\nu^\lambda \simeq {\rm diag.}(1,1,1-\Delta_\tau)Y_\nu$.}} 
\be
m_\nu^\lambda \propto  \pmatrix{1&0&0\cr
                              0&1&0\cr 
                              0&0&1-\Delta_{\tau}}
                     \,\,\,\, m_\nu\,\,\,\,
                     \pmatrix{1&0&0\cr
                              0&1&0\cr 
                              0&0&1-\Delta_{\tau}}.
\ee
\noindent
The proportionality involves a 
scale factor which is not  relevant to our present analysis.
The factor $\Delta_\tau$ is due to one loop RG evolution and we can neglect
$m_e^2$ and $m_\mu^2$ terms as compared to $m_\tau^2$. 
$\Delta_\tau$ is given approximately by 
\be
\Delta_\tau \simeq \frac{m_\tau^2}{8\pi^2 v^2}{(\tan^2\beta +1)}\ln\left(\frac{\Lambda}{\lambda}
\right),
\ee
\noindent
where $\tan\beta$ is the ratio of the VEVs of the up-type and down-type neutral Higgs 
fields in the MSSM and $v^2$ is twice the sum of their squares. 
Suppose the $\mu\tau$ symmetric form of $m_\nu$ is written as 
\be 
m_\nu = m\pmatrix{P&Q&Q\cr
                  Q&R&S\cr
                  Q&S&R},
\ee
where the complex quantities $P, Q, R, S$ 
are to be identified from the neutrino mass matrices 
given in eqn.(2.4) or (2.16). 
Then the  corresponding neutrino mass matrix at the low energy scale $\lambda$ comes out as 
\be
m_\nu^\lambda = m\pmatrix{P&Q&Q(1-\Delta_\tau)\cr
                            Q&R&S(1-\Delta_\tau)\cr
                      Q(1-\Delta_\tau)&Q(1-\Delta_\tau)&R(1-2\Delta_\tau)}.
\ee
\vskip 0.1in
\noindent
From eqn.(4.4) we can calculate $R^\lambda$ as well as 
 $\sin\ts12^\lambda$, $\sin\ta23^\lambda$ and $\sin\t13^\lambda$  
for the allowed textures of category A 
and category B. The corresponding analytic expressions are given in the 
Appendix.
There is now a slight extension of the allowed regions in the $k_1$-$k_2$ 
plane for category 
A and in the $l_1$-$l_2$ plane for category B are shown in 
Figs.3 and 4 respectively. 
For the allowed 
category A textures, we find that {\it any value of $\ta23^\lambda$ greater 
than $45^\circ$ is disallowed}. 
Then the experimentally allowed $3\sigma$ ranges 
$30.7^\circ\leq\ts12^\lambda\leq 39.2^\circ$, $36^\circ\leq\ta23^\lambda\leq45^\circ$ 
and the maximum allowed value $\simeq 60$ 
of $\tan\beta$ \cite{bookproy} restrict $\t13^\lambda$ to 
$0^\circ\leq\t13^\lambda\leq2.7^\circ$. 
Similarly, for the allowed category B textures, 
we find that {\it any value of 
$\ta23^\lambda$ less than $45^\circ$ is excluded}. For the $3\sigma$ 
allowed ranges $45^\circ\leq\ta23^\lambda\leq54^\circ$ and 
$30.7^\circ\leq\ts12^\lambda\leq39.2^\circ$, 
$\t13^\lambda$ is found to be in the interval 
$0^\circ\leq\t13^\lambda\leq0.85^\circ$. 
\begin{center}
\begin{figure}
\hskip 1.20cm
\psfig{figure=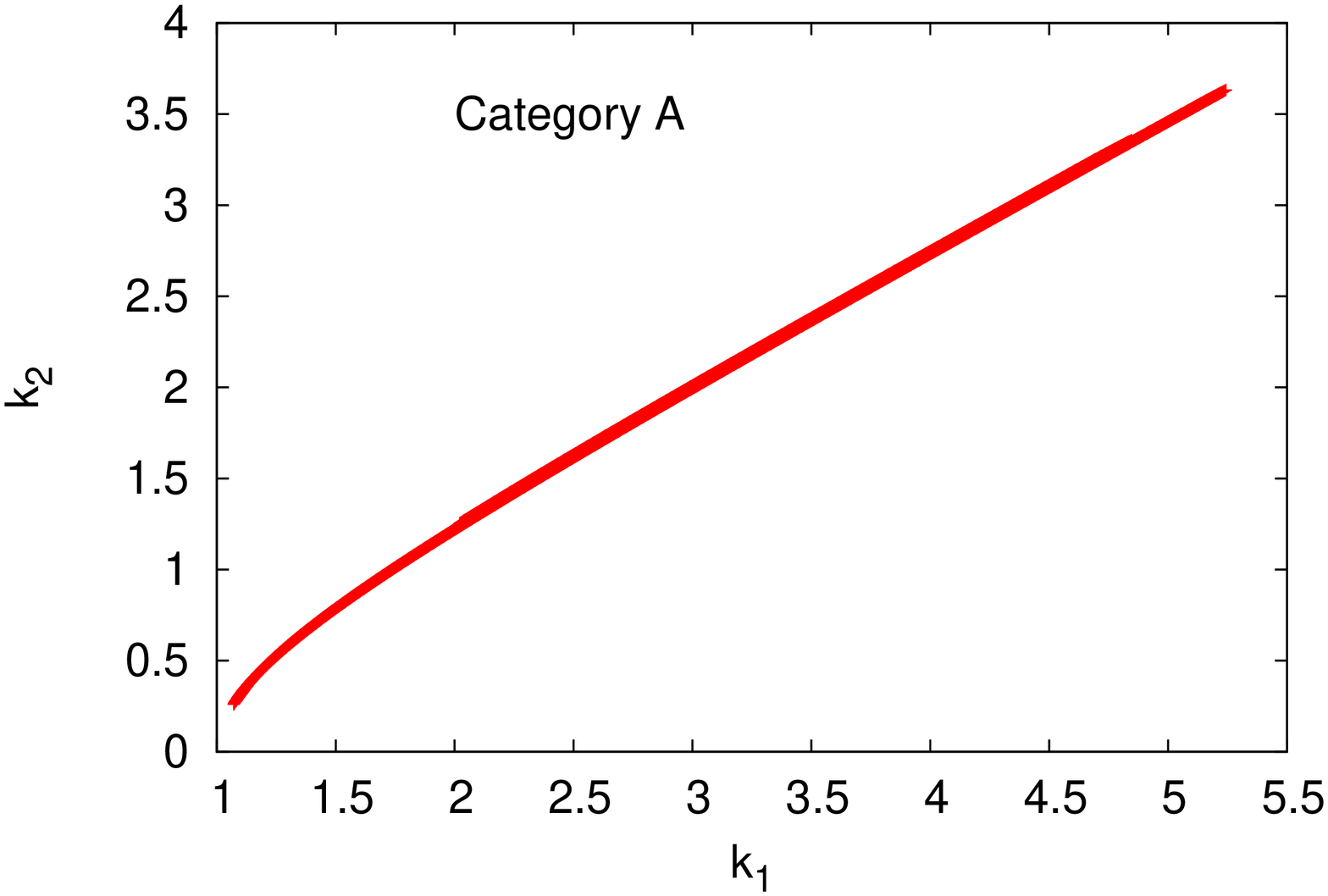,width=8.0cm,height=8.0cm,angle=270}
\caption{\small
The allowed variation of $k_1$ vs $k_2$ including radiative deviation 
within 3$\sigma$ allowed ranges of $R^\lambda$ and 
$\ts12^\lambda$. The phase angle $\alpha$ does not change significantly to $O(\Delta_\tau)$.}
\end{figure}
\end{center}
\begin{center}
\begin{figure}
\hskip 1.20cm
\psfig{figure=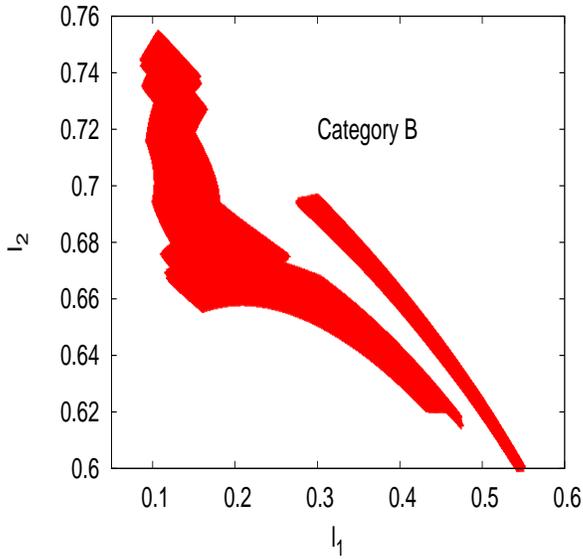,width=8.0cm,height=8.0cm,angle=270}
\caption{\small
The allowed variation of $l_1$ vs $l_2$
including radiative deviation  within 3$\sigma$ allowed ranges of 
$R^\lambda$ and 
$\ts12^\lambda$. The phase angle $\beta$ does not change significantly to $O(\Delta_\tau)$.
}
\end{figure}
\end{center}
\section{Concluding summary}
 This paper has investigated the effect of 
$\mu\tau$ symmetry and (more restrictively) TBM on the maximally 
allowed four zero neutrino Yukawa 
textures within the type I seesaw in the weak basis where 
charged leptons and the three heavy right 
chiral neutrino are mass diagonal. 
Only two textures (leading to the same from of $m_\nu$) out of 
fifty four in category A and two textures 
(again leading to an identical $m_\nu$ form) out 
of eighteen in category B survive the imposition 
of $\mu\tau$ symmetry. Each $m_\nu$ can be characterized by two 
real parameters and one phase: chosen to be $k_1$, $k_2$, $\alpha$
for category A and $l_1$, $l_2$, $\beta$ for category B. All are severely constrained by extant neutrino oscillation data. In each category, the additional requirement of TBM reduces the three parameters to a single real constant with a nearly fixed value.
\vskip 0.1in
\noindent
We have further looked at radiative lepton flavor violating decays 
$l_\alpha\rightarrow l_\beta\gamma$ (with $\alpha>\beta$=1,2,3) 
in the mSUGRA version of the MSSM. 
Our conclusion is that ${\rm BR}(\tau\rightarrow \mu\gamma)$ = 0 for 
category A and BR$(\tau\rightarrow e\gamma)$/BR$(\mu\rightarrow e\gamma)$\,\,
$\simeq$\,\, 0.178 
for both categories. 
Leptogenesis has also been considered at the energy scale min.($M_1$, $M_2$, $M_3$) with the following result. The term 
${\cal J}_{ij}^\alpha$, which does not contribute to the flavor-summed 
lepton asymmetry, vanishes in either category. 
The term ${\cal I}_{ij}^\alpha$, 
which can cause such an asymmetry, is constrained. 
In particular, (1) ${\cal I}^e_{12}$ and ${\cal I}^e_{13}$ are nonzero while 
the other contributions vanish in category A; (2) either ${\cal I}_{12}^\mu$, 
${\cal I}_{13}^\tau$ or ${\cal I}_{13}^\mu$, ${\cal I}_{12}^\tau$ are  
 nonzero with the rest vanishing in category B. Regarding effective washout masses, only $\tilde{m_1^e}$ is nonvanishing in category A, while all of  
$\tilde{m_1^e}$,  $\tilde{m_1^\mu}$,  $\tilde{m_1^\tau}$ 
are nonzero in category B with $\tilde{m_1^\mu}$ =  $\tilde{m_1^\tau}$.
\vskip 0.1in
\noindent
Finally, deviations from $\mu\tau$ symmetry, that are radiative in origin, 
have been considered. First, this symmetry has been imposed on $m_\nu$ at 
$\Lambda\sim {10}^{12}$ GeV which typifies an energy scale that is 
characteristic of the heavy right chiral neutrino masses. 
Then the deviations in the elements of $m_\nu$, 
caused by one-loop RG running from $\Lambda$ to the 
laboratory scale $\lambda\sim {10}^{3}$ GeV, have been 
computed in the MSSM with the largest allowed value of $\tan\beta$. 
Using the experimental $3\sigma$ ranges of $R$ and $\ts12$, 
we have found the following results : 
(1) category A allows only an inverted neutrino mass ordering 
$(\Delta m^2_{32}<0)$ with $\ta23\leq 45^\circ$ 
and $0^\circ\leq\t13\leq2.7^\circ$; 
(2) only a normal mass ordering $(\Delta m^2_{32}>0)$ with 
$\ta23\geq 45^\circ$ 
and $0^\circ\leq\t13\leq 0.85^\circ$ are allowed in category B. These 
predictions will face crucial future tests of the 
allowed four zero neutrino Yukawa textures 
in our scenario. Our bottom line is that $m_\nu^{(A)}$ is on the verge of exclusion, while $m_\nu^{(B)}$ is a good candidate for the true $m_\nu$ occurring in nature. A measured value of $\theta_{13}$ will provide a crucial test of the latter's viability.  \newpage
\appendix
\section{Expressions for measurable quantities}
{\bf $\mu\tau$ symmetric case}
\vskip 0.1in
\noindent
Eqn.(4.3) leads to 
\be
h = m_\nu m_\nu^\dagger = m^2\pmatrix{
|P|^2 +2|Q|^2& PQ^\star + Q(R^\star+S^\star)&  PQ^\star + Q(R^\star+S^\star)\cr
 P^\star Q + Q^\star(R+S)&|Q|^2+|R|^2+|S|^2&|Q|^2+RS^\star+R^\star S\cr
 P^\star Q + Q^\star(R+S)&|Q|^2+R^*S+RS^*&|Q|^2+|R|^2+|S|^2
}.
\ee
\noindent
The diagonalization of $h$ yields ${\rm diag.}{(m_1^2, m_2^2, m_3^2)}$ and also expressions for five relevant measurable quantities. The latter are : (1) $\Delta m_{21}^2 = m_2^2 - m_1^2$, i.e the light neutrino mass squared 
difference relevant to solar/reactor experiments, (2) the corresponding mixing angle $\ts12$, (3) $\Delta m_{32}^2 = m_3^2 - m_2^2$, i.e the neutrino mass squared difference pertaining to atmospheric/long-baseline studies, (4) the corresponding
mixing angle $\ta23$ and (5) the remaining mixing angle $\t13$. 
\vskip 0.1in
\noindent
The last five quantities can all be expressed in terms of three real functions 
$X_{1,2,3}$ of the complex quantities P, Q, R, S appearing in $m_\nu$. These 
are defined as 
\bea
X_1 &=& 2\sqrt{2}|PQ^\star + Q(R^\star + S^\star)|,\nonumber\\
X_2 &=& |R+S|^2 - |P|^2,\nonumber\\
X_3 &=& |R + S|^2 - |P|^2-4(|Q|^2 +RS^\star +R^\star S).
\eea
\noindent
We then have
\bea
\Delta m_{21}^2 &= & m^2{(X_1^2+X_2^2)}^{1/2},\\
\ts12 & = & \frac{1}{2}\tan^{-1}\frac{X_1}{X_2},\\
\Delta m_{32}^2 & = & \frac{m^2}{2}[X_3-{(X_1^2+X_2^2)}^{1/2}],\\
\ta23 & = & \frac{\pi}{4},\\
\t13 & = & 0.
\eea
\vskip 0.1in
\noindent
{\bf Case with RG- broken $\mu\tau$ symmetry}
\vskip 0.1in
\noindent
We work to one loop and ignore $O(\Delta_\tau^2)$ terms. Now from eqn.(4.4) one 
derives that
\be
m_\nu^\lambda {m_\nu^\lambda}^\dagger
=h - m^2\Delta_\tau \pmatrix{2|Q|^2 & 2QS^\star & PQ^\star+ QS^\star+ 
                             3QR^\star\cr
                             2Q^\star S & 2|S|^2 & |Q|^2 + RS^\star + 3R^\star S\cr
                             P^\star Q + Q^\star S + 3Q^\star R& |Q|^2 + R^\star S + 3R S^\star & 
 2(|Q|^2 + |S|^2) + 4|R|^2
}.
\ee
The cumbersome diagonalization of $m_\nu^\lambda {m_\nu^\lambda}^\dagger$ 
is avoidable since the algebra simplifies in the specific 
cases of category A and category B. Let us reintroduce 
$c_{12} = \cos\theta_{12}$, $s_{12} = \sin\theta_{12}$ where $\theta_{12}$
is given in eqn.(A.4). We now define five functions $F_{1,...5}$ 
in terms of 
$c_{12}$, $s_{12}$ and elements of the $m_\nu$ matrix $P$, $Q$, $R$ and $S$. The five functions $F$  
are 
\bea
F_1 &=& -\sqrt{2}c_{12}^2\frac{\left\lbrace P^*Q + 3Q^*(R+S)\right\rbrace\left\lbrace PQ^\star + Q(R^\star+S^\star)\right\rbrace }{|PQ^\star + Q(R^\star+S^\star)|} +
4c_{12}s_{12}\left| R + S\right|^2\nonumber\\\nonumber\\&&+
\sqrt{2}s_{12}^2\frac{\left\lbrace PQ^* + 3Q(R^*+S^*)\right\rbrace\left\lbrace P^*Q + Q^*(R+S)\right\rbrace }{|PQ^\star + Q(R^\star+S^\star)|} ,\nonumber\\\nonumber\\
F_2 &=& -\sqrt{2}c_{12}\frac{\left\lbrace P^*Q + Q^*(3R-S)\right\rbrace\left\lbrace PQ^\star + Q(R^\star+S^\star)\right\rbrace }{|PQ^\star + Q(R^\star+S^\star)|} +
2s_{12}\left( |Q|^2 +2|R|^2 + RS^* -SR^*\right) ,\nonumber\\\nonumber\\
F_3 & = & -4\sqrt{2}c_{12}s_{12}\frac{\left\lbrace |PQ^\star + Q(R^\star+S^\star)|^2+2|Q|^2|R+S|^2+ Q^2P^*(R^*+S^*)+ {Q^*}^2P(R+S)\right\rbrace}{|PQ^\star + Q(R^\star+S^\star)|} \nonumber\\\nonumber\\&&-4(c_{12}^2-s_{12}^2)\left| R + S\right|^2\nonumber\\\nonumber\\
F_4 & =& -\sqrt{2}s_{12}\frac{\left\lbrace P^*Q + Q^*(3R-S)\right\rbrace\left\lbrace PQ^\star + Q(R^\star+S^\star)\right\rbrace }{|PQ^\star + Q(R^\star+S^\star)|} - 2c_{12}\left( |Q|^2 +2|R|^2 + RS^* -SR^*\right),\nonumber\\\nonumber\\
F_5 & = & 2\sqrt{2}c_{12}s_{12}\frac{\left\lbrace |PQ^\star + Q(R^\star+S^\star)|^2+2|Q|^2|R+S|^2+ Q^2P^*(R^*+S^*)+ {Q^*}^2P(R+S)\right\rbrace}{|PQ^\star + Q(R^\star+S^\star)|} \nonumber\\\nonumber\\&&-4\left| R - S\right|^2+ 4s_{12}^2 |Q|^2 +4c_{12}^2\left\lbrace |Q|^2 + \left| R + S\right|^2\right\rbrace .
\eea
Thus $F_3$ and $F_5$ are real, while $F_1$, $F_2$ and $F_4$ are in general complex. 
We now list the changed values of the earlier mentioned five measurable 
quantities. 
\bea
(\Delta m_{21}^2)^\lambda & = & \Delta m_{21}^2 + \frac{1}{2}m^2F_3\Delta_\tau,\\
\theta_{12}^\lambda & = & \sin^{-1}|s_{12} + \frac{m^2c_{12}}{2\Delta m_{21}^2}F_1^\star\Delta_\tau|,\\
(\Delta m_{32}^2)^\lambda & = & \Delta m_{32}^2 + \frac{1}{2}m^2F_5\Delta_\tau,\\
\theta_{23}^\lambda & = & \sin^{-1}\left|\frac{1}{\sqrt 2} + 
\frac{\Delta_\tau}{2\sqrt 2} m^2
\left(\frac{s_{12}F_2^\star}{\Delta m_{21}^2 + \Delta m_{32}^2} 
- \frac{c_{12}F_4^\star}{\Delta m_{32}^2}\right)\right|,\\
\theta_{13}^\lambda & = &  
\frac{\Delta_\tau}{2\sqrt 2} m^2\left|
\frac{c_{12}F_2^\star}{\Delta m_{21}^2+\Delta m_{32}^2} 
- \frac{s_{12}F_4^\star}{\Delta m_{32}^2}\right|
\eea
Note that, upto order $\Delta_\tau$, we can write the changed value of the ratio $R$ 
as 
\be
R^\lambda = \frac{\Delta m_{21}^2}{\Delta m_{32}^2} 
+ \frac{1}{2}m^2\Delta_\tau
\left(
\frac{F_3}{\Delta m_{32}^2} - 
F_5\frac{\Delta m_{21}^2}{{(\Delta m_{32}^2)}^2}\right).
\ee
For convenience, we list the quantities $P$, $Q$, $R$ and $S$ in each category below: 
\vskip 0.1in
{\bf Category A}  From elements of $m_\nu^{(A)}$ in eqn.(\ref{mnuA})
\bea
P & = & k_1^2e^{2i\alpha} + 2k_2^2
,\nonumber\\
Q & = & k_2,\nonumber\\
R & = & 1,\nonumber\\
S &=& 0.
\eea
\vskip 0.1in
{\bf Category B}  From elements of $m_\nu^{(B)}$ in eqn.(\ref{mnuB})
\bea
P & = & l_1^2 ,\nonumber\\
Q & = & l_1l_2 e^{i\beta},\nonumber\\
R & = &l_2^2e^{2i\beta}+ 1,\nonumber\\
S & = & l_2^2e^{2i\beta}.
\eea

\end{document}